\documentclass[a4paper,12pt]{article}
\usepackage{graphicx}

\begin{document}

\title{Electroproduction of electron-positron pair in a medium}
\author{V. N. Baier
and V. M. Katkov\\
Budker Institute of Nuclear Physics,\\ Novosibirsk, 630090, Russia}

\maketitle

\begin{abstract}
The process of electron-positron pair creation by a high-energy
electron in a medium is analyzed. The spectral distribution over
energies of created particles is calculated for the direct and
cascade mechanisms of the process. The Coulomb corrections are
included. The new formulation of the equivalent photons method is
developed which takes into account the influence of multiple
scattering. It is shown the effects of multiple scattering can be
quite effectively studied in the process under consideration.
\end{abstract}

\newpage

\section{Introduction}
A high-energy electron passing through a medium produces the
electron-positron pair side by side with the radiation. The
electroproduction process possesses many important peculiarities
which will be discussed below.

Generally speaking one has to consider two mechanisms of the
process, The first one is the direct(one-step) electroproduction of
pair via the virtual intermediate photon. The second one is the
cascade(two-step) process when the electron emits the real photon at
collision with one nucleus which converts into the pair on another
nucleus. The interrelation of two mechanisms depends on the target
thickness $l$ since the probability of the direct process is
proportional to $l$ while the probability of the cascade process is
proportional to $l^2/2$. These contributions have comparable values
when the ratio $l/L_{rad}~(L_{rad}$ is the radiation length) is of
the order of a few percent.

For the direct process the differential, over the energies of
produced particles $\varepsilon_+,\varepsilon_-$, probability was
found by Kelner \cite{K} in the lowest over $Z\alpha$ ($Z$ is the
charge of a nucleus, $\alpha$=1/137) order of the perturbation
theory (for the derivation see also Sec.26 in \cite{BKF}). For the
direct process there are two types of diagrams: one-photon diagram
when the pair is created by one virtual photon emitted from the
electron at collision with a nucleus, and two-photon diagrams when
the pair is created by two virtual photons connected with the
initial electron and a nucleus. In the present paper the Coulomb
corrections are included in the probability contributions of both
types of diagrams.

In the soft part of created particles spectrum
($\varepsilon_-,\varepsilon_+ \ll \varepsilon,~\varepsilon$ is the
energy of the initial electron) one can neglect the one-photon
contribution and the two-photon contribution can be obtained using
the equivalent photons method within the logarithmic accuracy (see
Appendix B in \cite{BFKK}). At low enough energies
$\varepsilon_-,\varepsilon_+ $ the multiple scattering of the
initial electron results in distortion of the spectral distribution
of the equivalent photons which can lead to a modification of the
process probability. It is shown the effect of multiple scattering
can be quite effectively studied in electroproduction process. Just
as in the radiation process the effect can be observed in the soft
part of spectrum.

In Sec.2 the probabilities of the direct process are presented. The
new formulation of the equivalent photons method is given which
includes the influence of multiple scattering. This method permits
to find the alteration in the soft part of spectral distribution of
created particles. In Sec.3 the probabilities of the cascade process
taking into account the multiple scattering are considered.  The
probabilities, differential over one of created particle energy, are
analyzed in Sec.4

\section{Direct electroproduction probability with Coulomb corrections}

The Coulomb corrections to the direct electroproduction probability
can be found using the method outlined in the review \cite{BK} (see
also Appendix A in \cite{BK1}). The probability for the one-photon
diagrams contribution with the Coulomb corrections taken into
account in the case of complete screening has the form (the system
$\hbar=c=1$ is used)
\begin{eqnarray}
&& \frac{dw_1}{dzdy}=\frac{\alpha l}{\pi L}\frac{1-y}{y^2}
\Bigg\lbrace\left(A_1\ln(1+\xi)+B_1+C_1\frac{\xi}{1+\xi}\right)
\left(1+\frac{\ln(1+\frac{1}{\xi})}{2L_0}\right)
\nonumber \\
&& +\frac{1}{2L_0}\Bigg[A_1 {\rm Li}_2\left(\frac{\xi}{1+\xi}\right)
+B_1\frac{\ln(1+\xi)}{\xi}+ C_1\frac{2}{3}\frac{\xi}{1+\xi}
\nonumber \\
&& +\frac{2}{9}\left(\left(1-\beta\left(1+
\frac{1}{\xi}\right)\right)\ln(1+\xi)+\beta\right)\Bigg]\Bigg\rbrace,
 \label{1}
\end{eqnarray}
where $m$ is the electron mass, ${\rm Li}_2(x)=-\int_0^x
\frac{\ln(1-t)}{t}dt$ is the Euler dilogarithm,
\begin{eqnarray}
&&
z=\frac{\varepsilon_+}{\varepsilon},~y=\frac{\omega}{\varepsilon},
~\omega=\varepsilon_+ +\varepsilon_-,\quad \frac{1}{L}=\frac{4
Z^2\alpha^3n_aL_0}{m^2},\quad L =\L_{rad}
\left(1+\frac{1}{18L_0}\right),
\nonumber \\
&&L_0=\ln(183Z^{-1/3})-f(Z\alpha),\quad
f(x)=\sum_{n=1}^{\infty}\frac{x^2}{n(n^2+x^2)},
\nonumber \\
&& A_1=\xi\left(\frac{1}{2\beta}-1\right)-\frac{4}{3}\beta
\left(1+\frac{1}{\xi}\right)-\frac{2}{3},\quad
B_1=\frac{4}{3}\beta+\xi,\quad
\nonumber \\
&& C_1=\frac{4}{3}\beta+\frac{1}{3}\left(1+\xi\left(1-
\frac{1}{2\beta}\right)\right),\quad \beta=\frac{z(y-z)}{y^2},\quad
\xi=\frac{z(y-z)}{1-y},
 \label{2}
\end{eqnarray}
where $n_a$ is the number density of atoms in the medium,
$f(Z\alpha)$ is the Coulomb correction. It should be noted that in
Eq.(\ref{1}) at $\xi \ll 1$ a mutual compensation occurs in the
braces and the expression in the braces becomes proportional to
$\xi$.

The probability for the two-photon diagrams contribution with the
Coulomb corrections taken into account in the case of complete
screening has the form
\begin{eqnarray}
&& \frac{dw_2}{dzdy}=\frac{\alpha l}{\pi L}\frac{1-y}{y^2}
\Bigg\lbrace\left(A_2\ln(1+\frac{1}{\xi})+B_2+C_2\frac{1}{1+\xi}\right)
\left(1+\frac{\ln(1+\xi)}{2L_0}\right)
\nonumber \\
&& +\frac{1}{2L_0}\Bigg[A_2 {\rm Li}_2\left(\frac{1}{1+\xi}\right)
+B_2\xi\ln(1+\frac{1}{\xi})+ \frac{2}{3}\frac{C_2}{1+\xi}
\nonumber \\
&& -\frac{2}{9}\left(\left(\beta+\xi\left(1+\beta \right)
\right)\ln\left(1+\frac{1}{\xi}\right)-\beta\right)\Bigg]\Bigg\rbrace,
 \label{3}
\end{eqnarray}
where
\begin{eqnarray}
&& A_2=1-\frac{4}{3}\beta+ \xi\left(\frac{1}{2\beta}+
\frac{2}{3}(1-2\beta)\right),\quad B_2=\frac{4}{3}\beta-1,\quad
\nonumber \\
&& C_2=\frac{1}{3}\left( 4\beta-1-\xi \left(1+
\frac{1}{2\beta}\right) \right).
 \label{4}
\end{eqnarray}
The probabilities Eqs.(\ref{1}), (\ref{3}) are calculated within the
power accuracy (neglected terms $\propto m/\varepsilon_+,
m/\varepsilon_-$).

In the soft part ($y\sim z\ll1$) of spectral distribution $w_2$ one
can include the influence of multiple scattering on the initial
electron, the result is
\begin{equation}
\frac{dw_2^m}{dzdy}=\frac{\alpha l}{\pi
L}\frac{1-y}{y^2}\left[\Phi_2+\frac{\pi^2}{12L_0}\left(1-
\frac{4}{3}\beta\right)+\frac{1}{3}(4\beta-1)\left(1+
\frac{1}{3L_0}\right)\right],
 \label{5}
\end{equation}
where
\begin{eqnarray}
&&\Phi_2=P(y,z)\left(\ln\frac{1}{\xi}-1-\ln(1+\nu_1)\right). \quad
P(y,z)=1-\frac{4}{3}\beta\left(1+\frac{1}{12L_0}\right),
\nonumber \\
&&\nu_1=\frac{\varepsilon(1-y)}{\varepsilon_e y},\quad
\frac{m^2}{\varepsilon_e}=\frac{16\pi
Z^2\alpha^2n_aL_0}{m^2}=\frac{4\pi}{\alpha L},
 \label{6}
\end{eqnarray}
Here the function $\Phi_2$ describes the pair creation probability
in the equivalent photons method. Appearance of the term
$\ln(1+\nu_1)$ in the function $\Phi_2$ is connected with expansion
of the characteristic equivalent photon emission angles
$\vartheta_c$ due to the multiple scattering of the initial
electron. Let us consider this item in detail. The density of
equivalent photon can be presented as (see Eq.(B.7) in \cite{BFKK})
\begin{equation}
n(y,z)=\frac{\alpha}{\pi}\frac{1-y}{y}\left[\left(1+\frac{y^2}{2(1-y)}
\right)\ln \frac{Q^2(y,z)}{q_c^2}-1\right],
 \label{6a}
\end{equation}
where $Q^2=m^2 \omega^2/(\varepsilon_+\varepsilon_-)$ is the squared
minimal momentum transfer which is necessary for the photon with the
energy $\omega$ to create the pair with the energies $\varepsilon_+,
\varepsilon_-$. In absence of multiple scattering $q_c^2$ is defined
by the kinematics of the virtual photon emission from the initial
electron
\begin{equation}
q_c^2=q_{min}^2=\frac{m^2}{\varepsilon^2}\frac{\omega^2}{1-y},\quad
\frac{Q^2}{q_{min}^2}=\frac{\varepsilon^2(1-y)}{\varepsilon_+\varepsilon_-}
=\frac{1-y}{z(y-z)}=\frac{1}{\xi}.
 \label{6b}
\end{equation}
Taking into account the multiple scattering one has
$\varepsilon^2\vartheta_c^2/m^2=1+\nu_1$ (see e.g. Eqs.(2.10),
(2.25), (2.26) in \cite{BK}). So the equivalent photon spectrum at
$y\sim z \ll 1$ can be written as
\begin{equation}
q_c^2=\frac{\omega^2\vartheta_c^2}{1-y},\quad
n(y,z)dy=\frac{\alpha}{\pi}\frac{1-y}{y}\left[\ln \frac{1}{\xi}-1
-\ln(1+\nu_1) \right]dy.
 \label{6c}
\end{equation}
The differential probability of the pair creation by a photon in the
case of complete screening has the form
\begin{equation}
\frac{dw_p}{dz}=\frac{l}{L}\frac{P(y, z)}{y},
 \label{6d}
\end{equation}
and we find that the term $\propto \Phi_2$ in Eq.(\ref{5})
corresponds to the equivalent photon method. The contribution of the
multiple scattering is included in Eq.(\ref{6}) within the
logarithmic accuracy. Let us note that for heavy elements the value
$\varepsilon_e$ is of the order of a few TeV (e.g.
$\varepsilon_e$=2.73~TeV for tungsten, $\varepsilon_e$=2.27~TeV for
iridium), so for the electron energy of a few hundreds GeV, $\nu_1
\sim 1$ at $y \sim 1/10$.

\section{Cascade electroproduction probability}

It is known that the multiple scattering distorted the radiation
spectrum when $\nu_1\geq1$ or the photon energy
$\omega\leq\omega_c=\varepsilon^2/(\varepsilon+\varepsilon_e)$ (this
is the Landau-Pomeranchuk-Migdal(LPM) effect), while for the pair
creation process the LPM effect distorted the spectrum of created
pair only when $\omega\geq \omega_e=4\varepsilon_e$ (see \cite{BK}).
So for available electron energies one have to take into account the
multiple scattering in the cascade electroproduction process for the
radiation part only. Than the spectral distribution of cascade
process taking into account the multiple scattering of the initial
electron has the form
\begin{eqnarray}
&&\frac{dw_c}{dzdy}=\left(\frac{\alpha m^2 l}{4\pi
\varepsilon_e}\right)^2\frac{\varepsilon_e P(y,z)}{\varepsilon
y(1-y)} {\rm Im} \Bigg\lbrace y^2\left[\ln
p-\psi\left(p+\frac{1}{2}\right)\right]
\nonumber \\
&&+\left[2(1-y)+y^2\right] \left[\psi(p)-\ln p
+\frac{1}{2p}\right]\Bigg\rbrace,
 \label{7}
\end{eqnarray}
where $p=\sqrt{i}/(2\nu_1)$, $\psi(x)$ is the logarithmic derivative
of the gamma function (see Eq.(2.17) in \cite{BK}), the value
$\nu_1$ is defined in Eq.(\ref{6}). In the term describing radiation
(Im $\lbrace\rbrace$) the term  $\propto 1/L_0$ is neglected. The
contribution of this term doesn't exceed a few per cent under
considered conditions. This term is given by Eq.(2.33) in
\cite{BK1}) and can be included in the relevant case. In the case
when the multiple scattering of the initial electron may be
neglected ($\omega \gg \omega_c,~\nu_1 \ll 1,~|p| \gg1$) the
spectral distribution of cascade process is
\begin{equation}
\frac{dw_c^{QED}}{dzdy}=\frac{l^2}{L^2}\frac{P(y,z)}{2y^2}
\left[y^2+\frac{4}{3}(1-y)\left(1+\frac{1}{12L_0}\right)\right].
 \label{7a}
\end{equation}
Here the terms $\propto 1/L_0$, neglected in Eq.(\ref{7}), are taken
into account.

In the cascade process side by side with radiation inside of a
target one has to take into account the boundary radiation. Using
Eq.(3.12) in \cite{BK} (with regard for the factor 1/2 since photons
emitted at fly out of a target can't create the pair) and assuming
that $(\omega_0/(my))^2\ll 1+\nu_1$, where $\omega_0$ is the plasma
frequency (in any medium $\omega_0/m \leq 10^{-4}$) one has
\begin{equation}
\frac{dw_b}{dzdy}=\frac{\alpha^2 m^2 l}{4\pi^2 \varepsilon_e}
\frac{1-y}{y^2} P(y,z)\ln(1+\nu_1)
 \label{8}
\end{equation}
Putting together this probability and the probability Eq.(\ref{5})
we have that the terms with $\ln(1+\nu_1)$ are canceled. So the sum
of contributions of the equivalent and boundary photons doesn't
depend on multiple scattering.

At photon energy $\omega$ decreasing starting with $\omega=\omega_c$
($\nu_1=1$) the influence of multiple scattering on the radiation
process becomes significant. At this energy the estimation of the
interrelation of the different contributions is
\begin{equation}
\frac{dw_2^m+dw_b}{dw_c}= \frac{dw_2}{dw_c}\sim\frac{2L_{rad}}{l}
\frac{\alpha}{\pi} \ln\left(\frac{\varepsilon_e}{\varepsilon}\right)
 \label{9}
\end{equation}

With further decreasing of photon energy the relative contribution
of the cascade process is dropping both because of the logarithmic
growth of the probability $dw_2$  and because of the suppression of
the real photon emission probability due to the multiple scattering
of projectile (the LPM effect). In this interval of photon energies
(but for $\omega \gg m$) in the case when the value $l$ is low
enough the thickness of the target can become less than the photon
(virtual or real) formation length
\begin{equation}
l_f=\frac{2}{\omega\vartheta_c^2}=\frac{2\varepsilon^2}{m^2
\omega(1+\nu_1)},\quad \frac{l}{l_f}\simeq
\frac{l}{L_{rad}}\frac{2\pi(1+\nu_1)}{\alpha\nu_1^2}.
 \label{10}
\end{equation}
In this limiting case the contribution of real photons into the
process probability vanishes and only the contribution of virtual
photons remains. These photons build up outside a target where there
is no influence of the multiple scattering.

It should be noted that with photon energy decreasing the formation
length of pair creation by a photon inside target is also decreasing
($l_p=2\omega/m^2$) and we can be out of the complete screening
limit. However in this energy interval ($y \leq 10^{-3}$) the
equivalent photons method is applicable within the quite
satisfactory accuracy and the cross section of the photo-process is
known for arbitrary screening (see \cite{OM, BKF}).

\section{Partially integrated electroproduction probability}

The probability of electroproduction differential over one of
created particle energy only is of evident interest. It can be
obtained by integration of the found probabilities over $y~(z\leq
y\leq 1)$. For $z\ll1$ the main contribution into the integral gives
the region $y \sim z \ll 1$ (with the exception of the contribution
of one-photon diagrams which can be neglected in this energy
region).  For the ratio $r=dw_1/dw_2$ one has $r=0.011$ at $z=0.1$,
$r=0.042$ at $z=0.2$ and $r=0.24$ at $z=0.5$. Using Eq.(\ref{6}) at
$\nu_1=0$ and conserving the main term of decomposition over $z$ one
obtains for the summary contribution of the equivalent and boundary
photons
\begin{eqnarray}
&&\frac{dw_{b}}{dz}+\frac{dw_{2}^m}{dz}=\frac{dw_{2}}{dz}=\frac{2\alpha}{\pi
z}w_p \left(\ln\frac{1}{z}- \frac{1}{2} +\delta\right),
\nonumber \\
&& w_p=\frac{28}{9}\frac{Z^2\alpha^3}{m^2}l
n_a\left(L_0-\frac{1}{42}\right) \simeq \frac{7l}{9L_{rad}},\quad
\delta=\frac{\pi^2}{24L_0}-
\frac{1}{14}\left(1+\frac{1}{3L_0}\right).
 \label{11}
\end{eqnarray}
Here $w_p$ is the probability of pair creation by a photon in the
target with thickness $l$ in the case of complete screening
presented within power (relativistic) accuracy,neglected terms are
$\sim m/\varepsilon_+$. The quantity $\delta$ is the correction to
the equivalent photons method. This correction is small numerically:
e.g. for heavy elements ($L_0 \simeq 3.5,~\delta \simeq 1/25$), for
Ge $L_0 \simeq 4,~\delta \simeq 1/40$.

For the cascade process contribution in the region $z \ll
1,~\nu_1(z) \ll 1$ one gets
\begin{equation}
\frac{dw_{c}}{dz}=\frac{2l}{3L z}\left(1+ \frac{1}{12L_0}\right)w_p
\simeq \frac{14}{27z}\left(\frac{l}{L_{rad}}\right)^2.
 \label{12}
\end{equation}
The spectral distributions of created positrons reflect the spectral
distribution of photons (up to common factor $w_p$ in Eq.(\ref{11})
and $w_p/2$ in Eq.(\ref{12})).

When the parameter $\nu_1$ is large the asymptotic regime for the
radiation probability (see Eqs.(2.31),(2.32) of \cite{BK}) is
realized at very high value of $\nu_1$. Because of this one has to
use  Eq.(\ref{7}) directly. Conserving the main terms over $1/y$ we
find
\begin{equation}
\frac{dw_{c}}{dz} \simeq \left(\frac{l}{L_{rad}}\right)^2
\frac{2\varepsilon_e}{\varepsilon}\int_z^1
\left(1-\frac{4z(y-z)}{3y^2}\right){\rm Im} \left(\psi(x)-\ln x +
\frac{1}{2x}\right)\frac{dy}{y},
 \label{13}
\end{equation}
where $x=x(y)=\sqrt{i\varepsilon_e y/(4\varepsilon)}$.

Since the spectral distribution has the general factor $1/z$ its
characteristic properties at variation of $z$ over a few order of
magnitude one can track analyzing the function $z dw/dz$. For the
same reason in Figs 2 and 3 in the region under consideration ($z
\leq 0.2$) the dependence of this combination on the positron energy
$\varepsilon_+(z=\varepsilon_+/\varepsilon)$ is shown. The
probabilities Eqs.(\ref{3}), (\ref{7}) were used in calculation.

In the Fig.1 the spectral density $dw/dz$ for thin targets is shown.
The difference between curves 1 and 2 is due to the influence of
multiple scattering. Since the targets are quite thin, the
difference is still small especially for $l=170 \mu m$. The integral
$n_{12}=\int_{z_1}^{z_2} (dw/dz)~dz$ gives the number of positron
per one initial electron in the energy interval $\varepsilon
z_1-\varepsilon z_2$. For thickness $l=400~\mu m$ one has $n_{12}
\simeq 9\cdot 10^{-4}$ for the positron energies interval 0.5-5 GeV.
The targets of mentioned thicknesses were used in the experiment
NA63 carried out recently at SPS at CERN (for proposal see
\cite{NA63}).

The Fig.2 is another look on the process which permits to trace
details of the pair creation mechanism. The curves 2,5 in the right
part increase first tending to the asymptotic Eq.(\ref{12}) and than
decrease because of transition to the regime of Eq.(\ref{13}) at the
characteristic energy
$z_c=y_c=\omega_c/\varepsilon=\varepsilon/\varepsilon_e=0.01$. The
curves 1,4 are described nearly completely by Eq.(\ref{11}). The
increase of combination $z dw/dz$ is due to $\ln  1/z$. This
contribution dominates in the summary combination in the left part
of the spectrum for $l=400~\mu m$ and in the whole spectrum for
$l=170~\mu m$ (curves 3,6). Because of this the relative influence
of multiple scattering on the electroproduction process is falling.

The Fig.3 shows a different situation when the target is relatively
thick. Evidently here the influence of multiple scattering spreads
to the more wide positron energy interval (the region where the
curve 2 decreases). The cascade mechanism dominates for the positron
energy higher than 10 GeV.

In the Fig.4 the difference between the curves 1 and 2 shows the
influence of multiple scattering. This difference can be
characterized by ratio (see Eqs.(\ref{3}), (\ref{7}), (\ref{7a}))
\begin{equation}
\Delta=\frac{dw_{c}^{QED} - dw_{c}}{dw_{c}+ dw_{2}}.
 \label{14}
\end{equation}
In tungsten for the thickness $l=0.03~$cm one has at the initial
electron energy $\varepsilon=50~$GeV for the created positron energy
$\varepsilon_+=50$MeV ($z=0.001$) the value $\Delta \simeq$ 42\%,
and at the initial electron energy $\varepsilon=300~$GeV  for the
created positron energy $\varepsilon_+=100$ MeV the value $\Delta
\simeq$ 100\%.

\section{Conclusion}

The process of electron-positron pair production by a high-energy
electron traversing amorphous medium is investigated. It is shown
that the soft part of created particle spectrum may reduced due to
the multiple scattering of the initial electron. In the direct
process (via the virtual intermediate photon) the equivalent photon
spectrum is changed under the influence of multiple scattering. In
the cascade process (via real intermediate photon) the multiple
scattering distorted the photon spectrum inside a target. Besides,
the contribution of the boundary photons appears. It is shown,
within the logarithmic accuracy, that the change of the equivalent
photon spectrum is canceled by the contribution of the boundary
photons. As a result one has that the influence of multiple
scattering may be neglected in the very thin targets ($l \leq 1\%
L_{rad}$), where the direct process dominates in the soft part of
photon spectrum. The different situation arises in a more thick
target of heavy elements ($l \sim$ a few \% of $L_{rad}$). For the
initial electron energy in the range of hundreds GeV the multiple
scattering substantially diminish the spectrum of created positrons
in the range from hundreds MeV to a few GeV. This phenomenon can be
used for further study of the influence of multiple scattering on
higher order QED processes.

\vspace{0.5cm}

{\bf Acknowledgments}

The authors are indebted to the Russian Foundation for Basic
Research supported in part this research by Grant 06-02-16226.

\newpage

{\bf Figure captions}

{\bf Fig.1}

The summary spectral distribution $dw/dz=dw_2/dz+dw_c/dz$ of pair
electroproduction in amorphous germanium at the initial electron
energy $\varepsilon=180~$GeV. In the curves 1 and 3 (for two target
thicknesses $l=400~\mu m$ and $l=170~\mu m$ respectively) the
multiple scattering is taken into account (Eq.(\ref{7})), while in
the curves 2 and 4 the multiple scattering is neglected
(Eq.(\ref{7a})).

{\bf Fig.2}

The combination $z dw/dz$ for the pair electroproduction probability
in amorphous Ge at the initial electron energy
$\varepsilon=180~$GeV. The dotted curves 1 and 4 are the
contributions of two-photon diagrams Eq.(\ref{3}),  the dashed
curves 2 and 5 are the contributions of cascade process
Eq.(\ref{7}), the solid curves 3 and 6 are the sum of two previous
contributions for two thicknesses $l=400~\mu m$ and $l=170~\mu m$
respectively. For convenience the ordinate is multiplied by $10^3$.

{\bf Fig.3}

The combination $z dw/dz$ for the pair electroproduction probability
in amorphous tungsten at the initial electron energy
$\varepsilon=300~$GeV. The dotted curve 1 is the contribution of
two-photon diagrams Eq.(\ref{3}),  the dashed curve 2 is the
contribution of cascade process Eq.(\ref{3}), the solid curve 3 is
the sum of two previous contributions for the target thicknesses
$l=300~\mu m~(8.6\%~ L_{rad})$ For convenience the ordinate is
multiplied by $10^3$.

{\bf Fig.4}

The summary spectral distribution $dw/dz=dw_2/dz+dw_c/dz$ of the
pair electroproduction in amorphous tungsten of the thickness
$l=300~\mu m~(8.6\% L_{rad})$ at the initial electron energy
$\varepsilon=50~$GeV. In the curve 1 the multiple scattering is
taken into account (Eq.(\ref{7})), while in curve 2 the multiple
scattering is neglected (Eq.(\ref{7a})).


\newpage
\begin{thebibliography}{99}
\bibitem{K} S. R. Kelner, Yadernaya Fizica, {\bf 5} (1967) 1092.
\bibitem{BKF} V. N. Baier, V. M. Katkov, V. S. Fadin,
{\em Radiation from Relativistic Electrons} (in Russian) Atomizdat,
Moscow, 1973.
\bibitem{BFKK}V. N. Baier, V. S. Fadin, V.A.Khoze, E.A.Kuraev,
Phys. Rep. {\bf 78} (1981) 293.
\bibitem{BK}  V. N. Baier, V. M. Katkov,
Phys. Rep. {\bf 409} (2005) 261.
\bibitem{BK1}  V. N. Baier, V. M. Katkov,
Phys.Rev. {\bf D}57 (1998) 3146.
\bibitem{OM} H. Olsen, L. Maximon,
Phys.Rev. {\bf 114} (1959) 887.
\bibitem{NA63} J. U. Andersen, K.Kirsebom, S. P. Moller {\em et al},
{\em Electromagnetic Processes in Strong Cristalline Fields},
CERN-SPSC-2005-030.

\end{thebibliography}
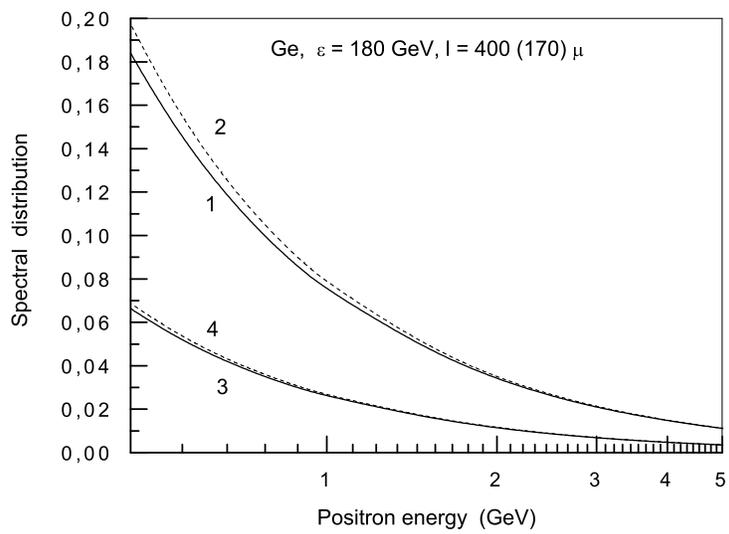
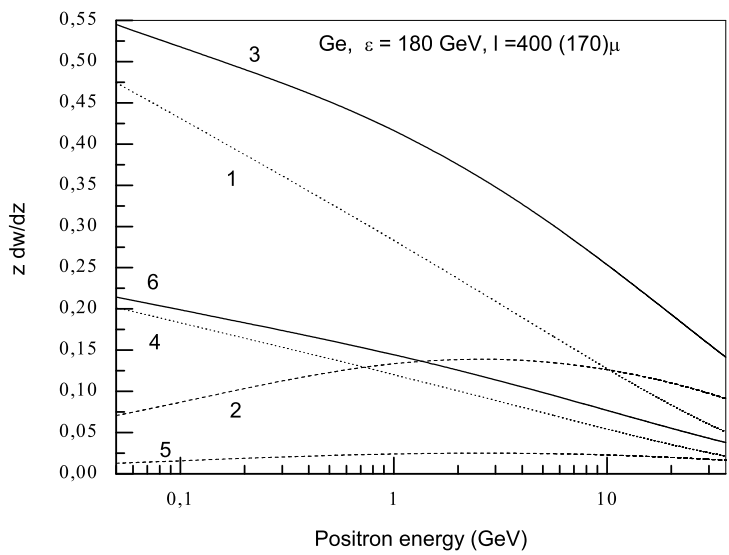
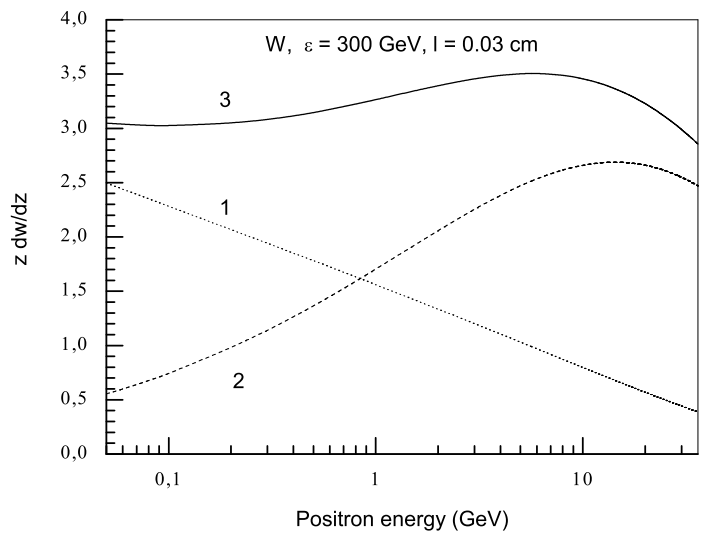
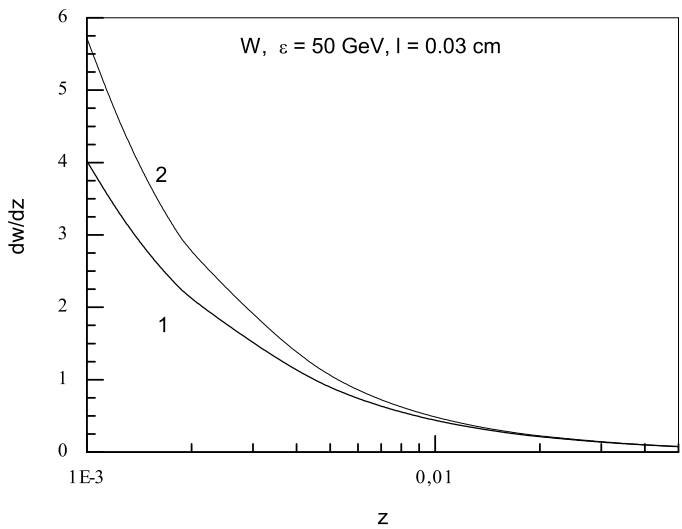
\end{document}